\documentclass[12pt]{article}

\usepackage{bm}
\usepackage{graphicx}
\usepackage{pdfsync}
\usepackage{graphicx}
\usepackage{hyperref}
\usepackage{amsmath}
\usepackage{mathtools}
\usepackage{amsthm}                
\usepackage{amssymb}
\usepackage{wrapfig}
\usepackage{amsfonts}
\usepackage[margin=1.0in]{geometry}
\usepackage{color}
\usepackage{multirow}
\usepackage{tabularx}
\usepackage{caption}
\usepackage{mathrsfs}
\usepackage[ruled]{algorithm2e}

\usepackage{booktabs,caption}
\usepackage[flushleft]{threeparttable}
\usepackage{tcolorbox}
\usepackage{setspace}
\usepackage{appendix}
\usepackage{float}
\usepackage{natbib}
\usepackage{subfigure}

\newtheorem{theorem}{Theorem}[section]
\newtheorem{lemma}[theorem]{Lemma}

\newcounter{assumption}[section]

\newcounter{remark}[section]

\newcounter{example}[section]

\usepackage[T1]{fontenc}
\usepackage[utf8]{inputenc}
\usepackage{authblk}

\def\T{\intercal}
\newcommand{\y}{\bm{y}}
\newcommand{\x}{\bm{x}}
\newcommand{\B}{\bm{\beta}}
\newcommand{\RR}{\mathbb{R}}
\newcommand{\X}{\mathbf{X}}
\newcommand{\W}{\mathbf{W}}
\newcommand{\HH}{\mathbf{H}}

\def\spacingset#1{\renewcommand{\baselinestretch}
  {#1}\small\normalsize} \spacingset{1} \allowdisplaybreaks

\begin{document}

\title{Modern Subsampling Methods for Large-Scale Least Squares Regression}
\author[1]{Tao Li}
\author[1,*]{Cheng Meng}
\affil[1]{Institute of Statistics and Big Data, Renmin University of China}
\affil[*]{Correspontding author: Cheng Meng, chengmeng@ruc.edu.cn}
\renewcommand*{\Affilfont}{\small\it} 
\renewcommand\Authands{ and } 

\date{}
\maketitle


\doublespacing



\begin{abstract}

Subsampling methods aim to select a subsample as a surrogate for the observed sample. 
As a powerful technique for large-scale data analysis, various subsampling methods are developed for more effective coefficient estimation and model prediction. 
This review presents some cutting-edge subsampling methods based on the large-scale least squares estimation. 
Two major families of subsampling methods are introduced, respectively, the randomized subsampling approach and the optimal subsampling approach.
The former aims to develop a more effective data-dependent sampling probability, while the latter aims to select a deterministic subsample in accordance with certain optimality criteria.
Real data examples are provided to compare these methods empirically, respecting both the estimation accuracy and the computing time.

\end{abstract}

\noindent%
\vfill



\section{Introduction}
During recent decades, the rapid development of science and technologies enables researchers to collect data with unprecedented sizes and complexities.
In the meanwhile, large-scale datasets are emerging in all fields of science and engineering, from academia to industry.
For example, Facebook has over 1.75 billion active users who contribute to nearly 350 million photos which are uploaded to Facebook daily \footnote{Source from \url{https://www.omnicoreagency.com/facebook-statistics/}}.
Consider Twitter, around 6,000 tweets are tweeted on Twitter in a second \footnote{Source from   \url{https://www.dsayce.com/social-media/tweets-day/}.}.
In addition, viewers spent around 15 billion hours (1,712,000 years, which is still rising) on YouTube in a month, and the videos being uploaded to YouTube are at the rate of 72 hours per minute \footnote{Source from  \url{https://omnimediafilm.com/blogs/the-power-of-video-marketing-and-how-to-use-it/37-staggering-video-marketing-statistics-for-2018/.}}.
These social media platforms are collecting and generating massive datasets with various types, such as text data, image data, and video data.
For another example, the European Bioinformatics Institute, one of the world’s largest biology-data repositories, stores nearly 160 petabytes of data and back-ups about genes, proteins, and small molecules. 
Moreover, such huge amount of genomics data almost doubles annually \citep{cook2019european}.

The large-scale datasets emerging from all fields provide researchers with unprecedented opportunities for data-driven decision-making and knowledge discoveries. 
Nevertheless, traditional statistical and machine learning algorithms may fail to analyze these data due to considerable computational burden in terms of both time and memory. 
The task of analyzing large-scale datasets calls for innovative, effective, and efficient methods or algorithms for addressing the new challenges due to the explosion of data.

According to \cite{doug20013d}, the big data challenges can be evaluated in three main aspects, including volume, velocity, and variety. 
Specifically, the volume is the size related to both the dimension and the number of observations, velocity is the interaction speed with the data, and the variety means various data structures.
In this article, the authors mainly discuss the first scenario with a focus on the case that the number of observations $n$ far exceeds the data dimension $p$.
To alleviate the computational burden caused by large $n$, there has been a large number of studies dedicated to developing engineering solutions.
These solutions include cloud computing, designing more powerful supercomputers, parallel computing, among others.
More details of these methods are provided in Section~2.

Despite the effectiveness of the engineering solutions, efficient statistical solutions are still in high demand, making big data analysis manageable on general-purpose personal computers.
The subsampling method is a powerful technique that can be used to achieve this goal. 
A subsampling problem can be described as follows: given a $p$-dimensional sample $\{\x_i\}_{i=1}^n$ generated from an unknown probability distribution, the goal is to take a subsample $\{\x^*_i\}_{i=1}^r$, $r\ll n$, as a surrogate for the original sample.
The selected subsample is then processed by down-streaming analysis for coefficient estimation, model prediction, and statistical inference.

The usage of subsampling method in large-scale data analysis has been extensively considered in various fields such as linear regression \citep{ma2015leveraging,meng2017effective,zhang2018statistical,ma2020asymptotic,derezinski2018leveraged,clarkson2019minimax}, generalized linear regression \citep{ai2018optimal,wang2018optimal,ai2020optimal,yu2020optimal}, streaming time series~\citep{xie2019online,li2019online}, large-scale matrix approximation \citep{williams2001using,wang2013improving,alaoui2015fast,altschuler2019massively,wang2019scalable}, nonparametric regression \citep{gu2002penalized,ma2015efficient,zhang2018statistical,meng2020more,sun2020asympirical}, among others.



\begin{figure}[H]\centering
        \includegraphics[scale = 0.6]{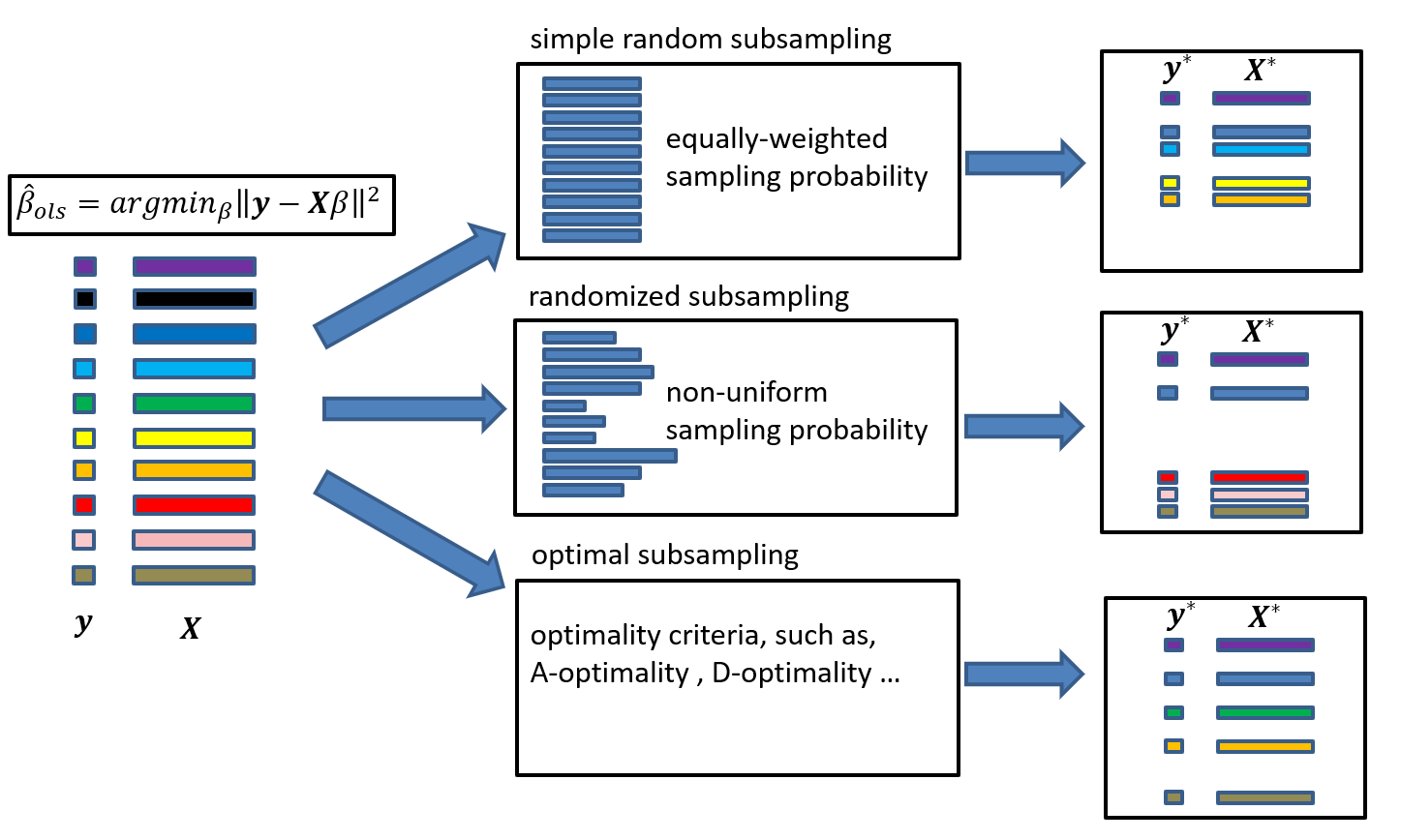} 
        \caption{Illustration for different subsampling methods. } \label{fig1}
\end{figure}

In this review, the authors mainly concentrate on the application of the subsampling approach in large-scale least squares regression.
One of the natural subsampling approaches is the simple random subsampling method, i.e., selecting a subsample with the equal-weighted sampling probability. 
Despite the simplicity, in numerous real-world applications, the simple random subsampling method may perform poorly, as the select subsample may not be an effective surrogate of the full sample \citep{cochran2007sampling,thompson2012simple}.
To overcome the limitation, there has been a large number of methods dedicated to developing more effective subsampling methods in the recent decade. 
Most of these methods can be divided into two classes, i.e., the randomized subsampling approach and the optimal subsampling approach, as illustrated in Fig.~\ref{fig1}. 

The randomized subsampling approach aims to improve the simple random subsampling method by carefully designing a data-dependent non-uniform sampling probability $\{\pi_i\}_{i=1}^n$, such that $\sum_{i=1}^n\pi_i=1$.
Intuitively, the data points that are more informative will be selected with a larger sampling weight.
One popular choice of the data-dependent sampling probability is the normalized statistical leverage scores, leading to the {\it algorithmic leveraging} approach \citep{ma2015statistical,meng2017effective,zhang2018statistical,ma2020asymptotic}.
After the subsample is selected, a weighted subsample least squares estimator will be calculated instead of a classic least squares estimator.
Different from the randomized subsampling approach, the optimal subsampling approach aims to construct a subsample estimator, which is most effective for coefficients estimation. 
In most cases, the subsample is selected based on certain rules, especially optimality criteria developed in the design of experiments, e.g., $A$-, $D$- and $E$-optimality \citep{pukelsheim2006optimal}.
A classic least squares estimator is then calculated based on the selected subsample.

The paper is structured as follows. 
Section~2 introduces some engineering solutions to tackle big data.
In addition, the authors also cover some other applications for subsampling methods in Section~2. 
Section~3 presents the essential background of the subsampling approach in the least squares regression problems.
Section~4 presents the details of the randomized subsampling approach and introduces several methods of this kind.
Section~5 outlines the basic concepts of optimal experimental design and the details of the optimal subsampling approaches.
Several real data examples are presented to compare the empirical performance of the subsampling methods, detailed in Section~6.
Section 7 summarizes the paper and discusses some open areas.

\section{Related works}

A large number of studies are dedicated to developing engineering solutions to alleviate the computational burden of big data.
These solutions include designing more powerful supercomputers, cloud computing, and parallel computing.
In particular, large quantities of supercomputers have been built rapidly in the past decade. 
The speed and storage of supercomputers can be hundreds or even thousands of times faster and larger in comparison with that of a general-purpose personal computer. 
However, the main limitation of supercomputers is that they consume enormous energy and may not be accessible to ordinary users.
Therefore, supercomputers are not a panacea for large-scale data analysis.

Cloud computing techniques can partially address the limitation of supercomputers and make computing facilities accessible to ordinary users. 
Nonetheless, the major bottleneck for cloud computing refers to that the inefficiency of transferring data due to the precious low-bandwidth internet uplinks, not to mention the problems of privacy and security concerns during the transfer process \citep{gai2012towards,yang2017big,stergiou2018security,sun2020security}.

The parallel computing technique, which is closely associated with the distributed algorithm and the divide and conquer method, solves big data problems in the following manner. 
The whole sample is first randomly partitioned into $k$ subsamples of equal sizes.
Each subsample is then analyzed by an independent local processor.
Finally, the results respecting all the local processors are combined to provide a final result by taking either a majority voting or a weighted average.
The parallel computing technique effectively alleviates the computational burden for big data analysis, respecting both time and memory.
Consequently, such a technique has been extensively studied in various statistical and machine learning fields \citep{brockwell2006parallel,bekkerman2011scaling,zhang2013divide, low2014graphlab,murray2016parallel,xu2018optimal}.
However, one limitation of parallel computing is that, to obtain a considerable reduction of the computational cost, one usually needs hundreds or thousands of local processors.
Such a large number of processors may not be obtainable by ordinary researchers.

Different from the aforementioned engineering solutions, the subsampling approach is a powerful statistical solution with the potential to make big data analysis manageable on general-purpose personal computers.
In addition, the subsampling approach has also been extensively applied in measurement constrained supervised learning, privacy-preserving analysis, efficient training of neural networks, and industry.

Measurement constrained supervised learning is an emerging problem in machine learning \citep{settles2012active,wang2017computationally,derezinski2018leveraged, clarkson2019minimax, meng2020lowcon}.
In this problem, the predictor observations are available, while the response observations are unavailable or are difficult to obtain.
Consider the task of predicting the soil functional property, i.e., the property related to a soil's capacity to support essential ecosystem service \citep{hengl2015mapping}. 
Suppose that a researcher wants to model the relationship between the soil functional property and some predictors, which can be derived from remote sensing data.
To obtain the response, i.e., the accurate measurement of the soil property, a sample of soil from the target area is required.
The response thus can be extremely time-consuming or even impractical to obtain, especially when the target area is off the beaten path. 
As a result, it is ideal to carefully identify a subsample of predictor observations, measure the corresponding responses, and then fit a supervised learning model based on the subsample of the predictors and responses.

The privacy-preserving analysis is taken as another example.
In some applications, the subsampling approach has the potential to enhance data security \citep{nissim2007smooth, li2012sampling}. 
Specifically, a carefully selected subset of data can reveal little confidential information \citep{shu2015privacy}.

Considering the training process of deep neural networks, it happens that data points are usually not equally important.
Specifically, some of the data points are properly handled after a few epochs of training, and others could be ignored at that point without impacting the final model.
Recently, researchers have shifted their focus on using importance sampling to improve and accelerate the training of neural networks \citep{bengio2009curriculum,alain2015variance,loshchilov2015online,schaul2015prioritized, katharopoulos2018not}.

In addition, subsampling methods also have wide applications in the industry. For example, subsampling methods help speed up the training process of statistical and machine learning models. 
Therefore, subsampling methods have the potential to benefit real-time monitoring systems, including abnormal detection systems, and fault diagnosis systems.
The authors refer to \cite{jiang2018recent,yin2019real,jiang2020review} for recent reviews of such areas.


\section{Problem formulation} 
Considering the classical linear model, 
\begin{eqnarray}\label{mmm1}
y_i=\x_i^\T\B_0+\epsilon_i, \quad i=1,2,\ldots,n,
\end{eqnarray}
where $y_i$'s are the responses, $\x_i$'s are the $p$-dimensional predictors ($p\ll n$), $\B_0\in \mathbb{R}^p$ is the vector of unknown coefficients, $\{\epsilon_i\}_{i=1}^n$ are i.i.d. random errors that follow a normal distribution with mean zero and constant variance $\sigma^2$, i.e., $\mathbb{N}(0,\sigma^2)$.
Let $\y\in\RR^n$ be the response vector and $\X\in\RR^{n\times p}$ be the predictor matrix.
The coefficient vector $\B_0$ can be estimated by calculating the ordinary least square (OLS), 
\begin{equation*}
    \widehat{\B}_{OLS}=\arg \min _{\B}\|\y-\X \B\|^2.
\end{equation*}
where $\|\cdot\|$ denotes the Euclidean norm. 
When $\X$ is full column rank, it is well-known that
\begin{equation*}
    \widehat{\B}_{OLS}=\left(\X^\T\ \X\right)^{-1} \X^\T \y.
\end{equation*}
Otherwise, when $\X$ is singular, $\left(\X^\T \X\right)^{-1}$ should be replaced by a generalized inverse of $\left(\X^\T \X\right)$. 
Let $\HH = \X(\X^\T \X)^{-1} \X^\T$.
The predicted response vector $\widehat{\y}$ can be calculated as
\begin{equation*}
    \widehat{\y}=\X\left(\X^\T \X\right)^{-1} \X^\T \y = \HH\y.
\end{equation*}
The matrix $\HH$ is thus often called the hat matrix, as it looks like a hat on response vector $\y$ to obtain $\widehat{\y}$. 
The hat matrix $\HH$ plays a crucial role in the randomized subsampling approach, which will be detailed in Section~4. 
To calculate the hat matrix $\HH$, people usually utilize the singular value decomposition (SVD) of $\X$ for robustness concern.
In particular, let $\X_{n \times p}=\mathbf{U}_{n \times n} \mathbf{\Lambda}_{n \times p} \mathbf{V}_{p \times p}^\T$ be the singular value decomposition of $\X$, where $\mathbf{U}$ and $\mathbf{V}$ are both orthonormal matrices and $\mathbf{\Lambda}$ is a diagonal matrix.
Through some calculations, it can be demonstrated that $\HH=\mathbf{U} \mathbf{U}^\T$ and $\widehat{\B}_{OLS}=(\X^\T \X)^{-1} \X^\T \y=\mathbf{V} \mathbf{\Lambda}^{-1} \mathbf{U}^\T \y$.

Although the least squares estimator $\widehat{\B}_{OLS}$ has a closed form, the computational cost for calculating the solution is of the order $O(np^2)$, which can be daunting when $n$ and/or $p$ are large.
To alleviate the computation burden, various subsampling methods are developed, from which the simple random subsampling may be the most natural one.
The simple random subsampling method works as follows.
Given a subsample size $r$, randomly select a subsample of size $r$ from the observed sample with equally weighted sampling probabilities, i.e., $\pi_i=1/n$, for $i=1,2,\dots,n$. 
Then, the least squares estimate is calculated purely based on the selected subsample.
Details are summarized as follows.

\begin{algorithm}
\caption{Simple random subsampling method}
\begin{tabbing}
\\
\textbf{Input:} $\mathbf{X}\in \RR^{n\times p}$, $\y\in \RR^n$, subsample size $r$\\
\quad Randomly sample $r$ observations using equally weighted sampling probabilities.\\
\quad Let $\y^*$ and $\X^*$ be the selected response vector and the matrix, respectively.\\
\textbf{Output:} $\widetilde{\B}_{RAND}=\arg \min _{\B}\|\y^*-\X^* \B\|^2$.
\end{tabbing}\label{algo1}
\end{algorithm}


Calculating the estimator $\widetilde{\B}_{RAND}$ requires only $O(rp^2)$ computation time, which is a significant reduction in comparison with $O(np^2)$, especially when $r\ll n$. 
In addition to the computational benefits, such an estimator also enjoys some theoretical benefits.
In particular, \cite{ma2015statistical} showed $\widetilde{\B}_{RAND}$ is an unbiased estimator respecting the true coefficient $\B_0$.
In addition, conditional on the sample, $\widetilde{\B}_{RAND}$ is unbiased to the full sample least squares estimator $\widehat{\B}_{OLS}$. 
Despite the algorithmic and theoretical benefits, Algorithm~\ref{algo1} may suffer from high estimation variance. 
Specifically, it can be shown that the estimation variance of $\widetilde{\B}_{RAND}$ is proportional to $n/r$ \citep{ma2015statistical}.
Consequently, in practice, when $r$ is relatively small compared to $n$, the simple random subsampling method may generate unacceptable results.

\begin{figure}[H]
        \centering
        \subfigure{
		  \includegraphics[scale=0.25]{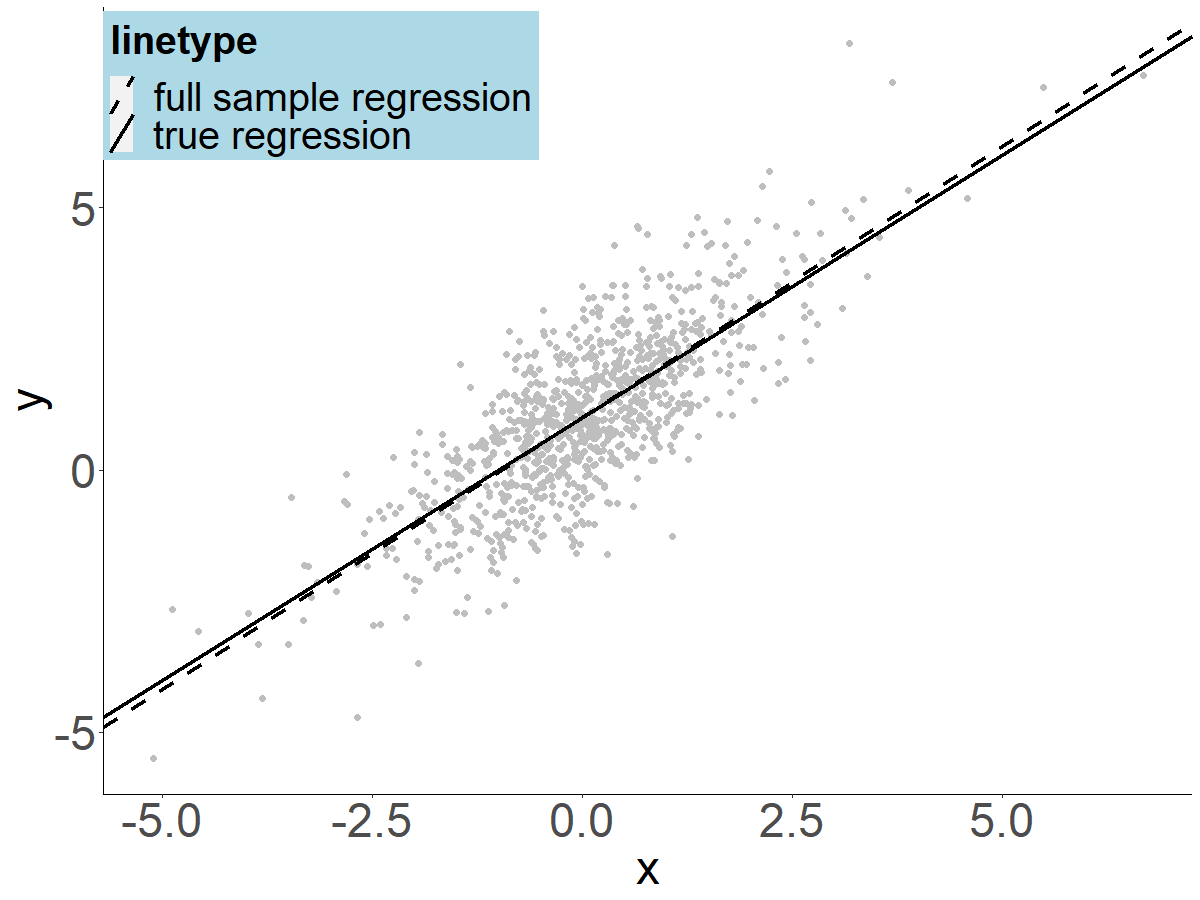}}
		\hspace{0.01in} 
		\subfigure{
		  \includegraphics[scale=0.25]{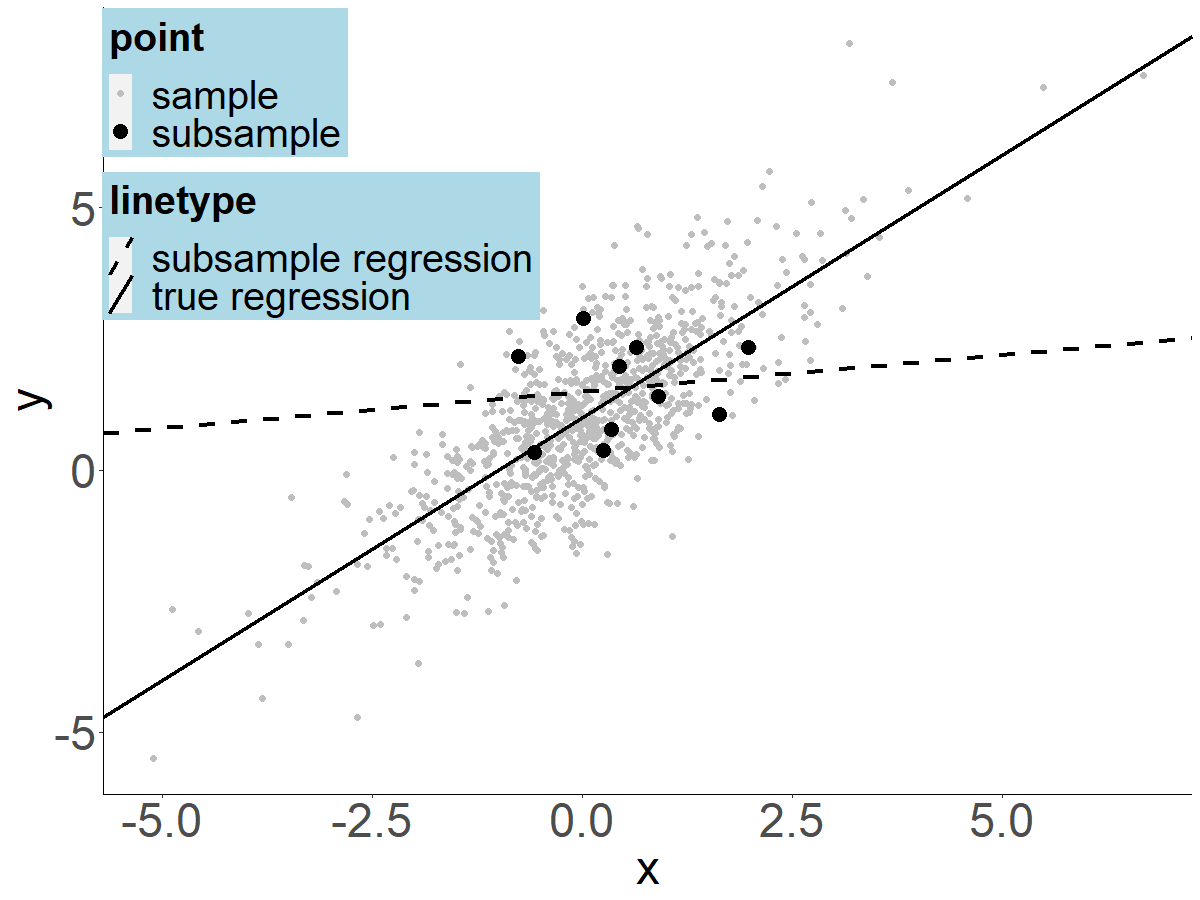}}
        \caption{Illustration of the unacceptable result of simple random subsampling method in least squares estimation. } 
        \label{fig2} 
\end{figure}

Figure~\ref{fig2} shows a toy example that Algorithm~1 fails to provide a decent estimate.
In the current example, data are generated from the model $y_i=x_i+1+\epsilon_i$, $i=1,\ldots,1000$, where $x_i$ is generated from $t$-distribution with 5 degrees of freedom and $\{\epsilon_i\}_{i=1}^n$ are the i.i.d. standard normal errors.
The data points are labeled as gray points, and the true regression line is labeled as the solid black line.
The left panel of Fig.~\ref{fig2} shows the full sample regression line, which is labeled as the dashed black line.
The authors observe that the full sample regression line can effectively estimate the true regression line.
Then the authors use the simple random subsampling algorithm to select a subsample of size ten, and label the selected data points as black dots in the right panel of Fig.~\ref{fig2}.
The dashed black line represents the fitted regression line of the selected subsample.
Obviously, the subsample linear regression line deviates severely from either the full sample regression line or the true regression line. 
Such an observation suggests that the performance of the simple random subsampling algorithm may deteriorate significantly when the subsample size $r$ is small.

To overcome the limitation of the simple random subsampling, existing methods proposed to utilize the bagging technique or the divide and conquer strategy \citep{chen2009bagging,lin2011aggregated,li2013statistical,zhang2013divide,chen2014split,xu2016feasibility}.
In particular, these methods calculate the subsample least squares estimator multiple times, and each estimator is calculated based on an independent random subsample.
Then, all the estimators are aggregate together by taking a weighted average to build the final estimator.
Existing literature shows the bagging techniques and the divide and conquers techniques can effectively reduce the variance for coefficients estimation \citep{buhlmann2002analyzing,li2013statistical,zhang2013divide}.
As a result, these techniques are extensively applied as a surrogate to the simple random subsampling method.
Despite the effectiveness, one limitation of these techniques refers to that they require a much larger number of response values in comparison with the simple random subsampling method does.
Thus, these techniques are not suitable for the measurement-constrained supervised learning or the privacy-preserving analysis, in which cases only a few numbers of the response values are available.

In the recent decade, a large number of studies are dedicated to developing more elegant subsampling methods, which select the subsamples that are more effective for large-scale least squares estimation.
These methods aim to develop a subsample least squares estimator that has a smaller estimation mean squared error (MSE) than the simple random subsampling estimator.
In addition, the proposed methods only require a relatively small number of response values.
These methods thus are suitable for both measurement-constrained supervised learning and the privacy-preserving analysis.

\section{Randomized subsampling methods}

In this section, the authors introduce a family of randomized subsampling methods to approximate the full sample least squares estimation. 
The randomized subsampling methods utilize data-dependent non-uniform sampling probability to select a random subsample, and then calculate the estimate based on such an example.
Such a weighted subsampling approach has long been studied in the literature of importance sampling to improve the numerical efficiency of the Monte Carlo approach \citep{clark1961importance,siegmund1976importance,glynn1996importance,liu2008monte}.
Algorithm~\ref{algo2} summarizes the general framework of the randomized subsampling method for least squares estimation.

\begin{algorithm}
\caption{General randomized subsampling method}
\begin{tabbing}
\\
\textbf{Input:} $\mathbf{X}\in \RR^{n\times p}$, $\y\in \RR^n$, subsample size $r$, sampling probability $\{\pi_i\}_{i=1}^n$\\
\quad Randomly sample $r$ observations using $\{\pi_i\}_{i=1}^n$.\\
\quad Let $\y^*$ and $\X^*$ be the selected response vector and the matrix, respectively.\\
\quad Denote $\{\pi_1^*,\ldots, \pi_r^*\}$ as the sampling probability respecting the selected observations.\\
\quad Let the diagonal matrix $\W=\mbox{diag}(1/\pi_1^*, \ldots, 1/\pi_r^*).$\\
\textbf{Output:} $\widetilde{\B}_{W}=\arg \min_{\B}(\y^*-\X^* \B)^\T \W (\y^*-\X^* \B)$.
\end{tabbing}\label{algo2}
\end{algorithm}


Algorithm~2 calculates a weighted least squares estimate instead of an ordinary least squares estimate in Algorithm~1. 
Note that when the sampling probability $\{\pi_i\}_{i=1}^n$ is equal-weighted, the ordinary least squares estimation and the weighted least squares estimation yield the same result.
However, for a general non-equally weighted sampling probability, \cite{ma2015statistical} showed that calculating the weighted least squares estimation is essential to obtain the estimator $\widetilde{\B}_{W}$ that is asymptotically unbiased to the full sample estimator $\widetilde{\B}_{OLS}$, condition on the sample.
Furthermore, \cite{ma2015statistical} showed that $\widetilde{\B}_{W}$ is asymptotically unbiased to the true coefficients $\B_0$.

Despite the theoretical effectiveness of Algorithm~\ref{algo2}, one fundamental question remains unanswered, i.e., how to decide the sampling probability $\{\pi_i\}_{i=1}^n$.
Intuitively, a good strategy for determining $\{\pi_i\}_{i=1}^n$ is to assign larger weights to the data points that are more "informative," i.e., the ones that are more influential to coefficients estimation and model prediction.
Furthermore, another fundamental question is that does there exist any sampling probability that is in some sense "optimal" respecting any optimality criterion.
To answer these questions, many randomized subsampling algorithms are developed in the recent decade.
Recall that $\HH=\X(\X^\T\X)^{-1}\X^\T$ denotes the hat matrix.
Let $h_{ii} = \x_i^\T(\X^\T\X)^{-1}\x_i$ be the $i$-th diagonal element of the hat matrix $\HH$.
Table~\ref{table1} summarizes some representative subsampling strategies with the proposed sampling probability $\{\pi_i\}_{i=1}^n$.
Table~\ref{table1} also summarizes some volume subsampling approaches, which utilizes a joint sampling distribution over subsamples to select a subsample directly, see \cite{derezinski2017unbiased} and \cite{derezinski2018leveraged} for more details.

\begin{table}
  \caption{Some representative randomized subsampling methods }
  \centering 
  \footnotesize
  \begin{threeparttable}
    \begin{tabular}{ccc}
    Methods  & Sampling Probability  & References\\
     \midrule\midrule
Simple Random Subsampling  &   
$\pi_i^{RAND}=1/n$     &   /                        \\
    \cmidrule(l  r ){1-3}
     Basic Leverage Subsampling & $ \pi_i^{BLEV} = h_{ii}/p$ & \cite{drineas2006sampling} \\ 
    \cmidrule(l r ){1-3}
     Shrinkage Leverage Subsampling & $\pi_i^{SLEV} = \alpha  h_{ii}/p +(1-\alpha)/n,\quad  \alpha\in[0,1] $ & \cite{ma2015statistical} \\ 
    \cmidrule(l r ){1-3}
     Inverse Covariance Subsampling & $\pi_i^{IC} = ||(\X^\T\X)^{-1}\x_i||/(\sum_{i=1}^n ||(\X^\T\X)^{-1}\x_i||)$ & \cite{ma2020asymptotic} \\ 
    \cmidrule(l  r ){1-3}
    Root Leverage Subsampling
    &  $\pi_i^{RL} =\sqrt{h_{ii}}/(\sum_{i=1}^n \sqrt{h_{ii}})$       & \cite{ma2020asymptotic}                      \\
    \cmidrule(l  r ){1-3}
     Predictor Length Subsampling & $\pi_i^{PL} = ||\x_i||/(\sum_{i=1}^n ||\x_i||)$ & \cite{ma2020asymptotic} \\

    \cmidrule(l  r ){1-3}
      Standard Volume Subsampling & $\operatorname{Pr}(S)=\frac{\operatorname{det}\left(\X_{S}^{\top} \X_{S}\right)}{\left(\begin{array}{c}n-p \\ r-p\end{array}\right) \operatorname{det}\left(\X^{\top} \X\right)}$ \tnote{*}& \cite{derezinski2017unbiased} \\
      
     \cmidrule(l  r ){1-3}
      Leveraged Volume Subsampling & 
      $\operatorname{Pr}(\tau) \propto \operatorname{det}\left(\sum_{i=1}^{r} \frac{1}{q_{\tau_{i}}} \x_{\tau_{i}} \x_{\tau_{i}}^{\top}\right) \prod_{i=1}^{r} q_{\tau_{i}}$ \tnote{**}& \cite{derezinski2018leveraged} \\
    \midrule\midrule
    \end{tabular}
    \begin{tablenotes}
    \item[*] where $S$ is the subset of $\{1,2,\ldots,n\}$ , $r$ is the size of $S$ and $\X_{S}$ is the submatrix of the rows from $\X$ indexed by the set $S$.
    \item[**] where $\tau \in \{1,2,\ldots,n\}^r$, $r>4p^2$ and $q_i=h_{ii}/p,i = 1,2,\ldots,n$.
    \end{tablenotes}
\end{threeparttable}
\label{table1}
  \end{table}

\subsection{Leverage scores and the basic leverage subsampling method}

Recall that the key to the success of the randomized subsampling algorithm is to carefully design a sampling probability,  reflecting the importance of each data point.
To achieve the goal, a metric is needed, which can quantitatively evaluate the importance of each data point.
In the statistical literature for model diagnostics, there exists the concept of leverage score to achieve this goal \citep{Weisberg2005}. 
For the $i$-th data point, its leverage score is defined as $\partial \widehat{y}_{i}/\partial y_i.$
Intuitively, if the leverage score is large, it indicates that a small disturbance in $y_{i}$ results in a big change in $\widehat{y}_{i},$ thus exerting a crucial role in model diagnostics.

Although the leverage score is defined upon the response $y_i$ and its fitted value, one important property of the leverage score is that, in the linear regression model, the leverage score of a data point is irrelevant to its response value and only associates with its predictors.
Specifically, recall that one has $\widehat{y_i} = \HH y_i$ in linear models.
Let $h_{ii} = \x_i^T(\X^\T\X)^{-1}\x_i$ be the $i$-th diagonal element of the hat matrix $\HH$, one thus has

\begin{equation}\label{eqn2}
\frac{\partial \widehat{y}_{i}}{\partial y_{i}}=\frac{\partial\left(\sum_{j=1}^{n} h_{i j} y_{j}\right)}{\partial y_{i}}=h_{i i},    
\end{equation}
i.e., the leverage score of the $i$-th data point equals $h_{ii}$, and is independent of the response~$y_i$.

Considering $e_i$, which is the residue respecting the $i$-th data point, it can be shown that $0<h_{ii}<1$ and
\begin{equation}\label{eqn3}
\mbox{Var}(e_i)=\mbox{Var}(\widehat{y}_i-y_i)=(1-h_{i i}) \sigma^2.
\end{equation}
Equation~(\ref{eqn3}) indicates that the data point with a high leverage score is associate with a small variance of the residue.
In other words, the linear regression line tends to pass close to these data points with high leverage scores, e.g., these points have a large impact on the linear regression line. 
When $p=1$, it can be shown that
$h_{i i}=1/n+\left(x_{i}-\bar{x}\right)^{2}/{\sum_{j=1}^{n}(x_{j}-\bar{x})^2}$, where $\bar{x}=\sum_{j=1}^n x_j/n.$
In such cases, the data points with large leverage
scores are the data points that are far away from the sample mean.

The basic leverage subsampling method (BLEV), also called \textit{Algorithmic Leveraging}, utilizes a sampling probability that is proportional to leverage scores, i.e., $\pi_i^{BLEV}\propto h_{ii}$, $i=1,\ldots,n$, \citep{drineas2006fast,ma2015statistical,ma2015leveraging,meng2017effective}.
Considering the sum of the leverage scores, when $\X$ has full column rank, one has
\begin{equation}\label{eqn4}
  \sum_{i=1}^n h_{i i}=\mbox{tr}(\HH)=\mbox{tr}\left(\X\left(\X^\T \X\right)^{-1} \X^\T\right)=\mbox{tr}\left(\left(\X^\T \X\right)^{-1} \X^\T \X\right)=\mbox{tr}\left(\mathbf{I}_{p}\right)=p.
\end{equation}
Equation~(\ref{eqn4}) indicates that the BLEV sampling probability
\begin{equation*}
    \pi_i^{BLEV}=\frac{h_{ii}}{p} =\frac{\x_i^\T(\X^\T\X)^{-1}\x_i}{p}, \quad i=1,\ldots,n,
\end{equation*}
is valid, since $0<h_{i i}/p<1$ and $\sum_{i=1}^{n} (h_{ii}/p)=1.$

The computational cost for the BLEV method mainly incurs in calculating the leverage scores.
To calculate the leverage scores, one can utilize the singular value decomposition of the observed sample, resulting in a computational cost of the order $O(np^2)$.
However, such a computational cost is the same as the one for calculating the full sample least squares estimation.
Fortunately, there are random projection-based techniques for approximating the leverage scores efficiently, consequently reducing the running time from $O(np^2)$ to $o\left(n p^2\right)$ \citep{drineas2012fast}.
The authors refer to \cite{avron2010blendenpik,meng2014lsrn,gittens2013revisiting} for further development of such techniques.

The basic leverage subsampling method has long been studied and has been extensively applied in various fields \citep{papailiopoulos2014provable,rudi2018fast,xie2019online,agarwal2020leverage}.
A variety of empirical studies show such a method outperforms the simple random subsampling method \citep{drineas2006sampling,mahoney2011randomized,drineas2012fast}.
In particular, \cite{drineas2006sampling} adopted an algorithmic perspective that concentrates on providing worst-case run-time bounds for different inputs. 
The authors showed that the basic leverage subsampling method provides worst-case algorithm results that are uniformly superior to the simple random subsampling method.
Despite the effectiveness of BLEV, some natural questions remain unanswered.
For example, whether the basic leverage sampling probability $\{h_{ii}/p\}_{i=1}^n$ can be improved?
Moreover, what is an appropriate metric to evaluate the effectiveness of a sampling probability?
In particular, how to develop a sampling probability that is optimal, respecting such a metric?
To answer these questions, a large number of sampling probabilities are developed in the recent decade, which are detailed as follows.

\subsection{Advanced leverage subsampling methods}

One natural criterion to evaluate different randomized subsampling methods is to quantify the difference between the subsample estimates and the true coefficients $\B_0$.
To achieve the goal, \cite{ma2020asymptotic} measured such a difference based on the asymptotic mean squared error (AMSE) respecting the true coefficients $\B_0$ and developed the following lemma for a general randomized subsample estimator $\widetilde{\B}_W$, which was introduced in Algorithm~\ref{algo2}.

\begin{lemma}\label{thm1}
Assume the number of predictors $p$ is fixed and the following regularity conditions hold.
\\
\textbf{Condition (1):} There exist positive constants $b$
and $B$ such that $b \leq \lambda_{\min } \leq \lambda_{\max } \leq B,$ where $\lambda_{\max }$
and $\lambda_{\text {min}}$ are the maximum and minimum eigenvalues of matrix $\mathbf{X}^{T} \mathbf{X} / n$, respectively. 
\\
\textbf{Condition (2):} The subsample size satisfies $r=O\left(n^{1-\delta}\right)$, $0 \leq \delta<1$. The minimum sampling probability $\pi_{min }=O\left(n^{-\gamma_{0}}\right),$ where $\gamma_{0} \geq 1$ and $\gamma_{0}+\alpha<2$.
\\
Under Condition (1)-(2), as $n \rightarrow \infty$, one has
\begin{equation*}
    \left(\sigma^{2} \bm{\Sigma}_{0}\right)^{-\frac{1}{2}}\left(\widetilde{\B}_W-\B_0\right) \stackrel{d}{\rightarrow} N\left(\mathbf{0}, \mathbf{I}_{p}\right)
\end{equation*}
Here, $\bm{\Sigma}_0=(\X^\T \X)^{-1}(\X^\T(\mathbf{I}_n+\bm{\Omega}) \X)(\X^\T \X)^{-1}$, where $\bm{\Omega}=\mbox{diag}\{1 / r \pi_{i}\}_{i=1}^{n}$ and $\mathbf{I}_{p}$ is the $p \times p$ identity matrix.
\end{lemma}

As shown in Lemma~\ref{thm1}, Condition (1) indicates that the matrix $\X^\T\X$ is positive definite, i.e., the matrix $\X$ has full column rank.
Condition (2) provides a lower bound on the smallest sampling probability.
In addition, Condition (2) suggests that bounding sampling probabilities from below mitigates the inflation of the estimation variance.
When Condition (1)-(2) holds, Lemma~\ref{thm1} indicates that the general randomized subsample estimator $\widetilde{\B}_W$ is an unbiased estimator to the true coefficients $\B_0$.
Furthermore, the AMSE of $\widetilde{\B}_W$ is equivalent to its asymptotic variance, which can be written as
\begin{equation}\label{eqn5}
\mbox{AVar}(\widetilde{\B}_W) = \sigma^2(\X^\T\X)^{-1}+\sigma^2(\X^\T\X)^{-1}\X^\T\bm{\Omega}\X(\X^\T\X)^{-1}.
\end{equation}
Equation~(\ref{eqn5}) indicates that the asymptotic variance of the estimator $\widetilde{\B}_W$ includes two parts, where the first part $\sigma^2(\X^\T\X)^{-1}$ is the variance of the full sample OLS estimator, and the second part $\sigma^2(\X^\T\X)^{-1}\X^\T\bm{\Omega}\X(\X^\T\X)^{-1}$ is associated with the subsampling procedure.
To obtain the optimal sampling probability respecting the AMSE, one thus only needs to find the $\bm{\Omega}$ that can minimize $(\X^\T\X)^{-1}\X^\T\bm{\Omega}\X(\X^\T\X)^{-1}$.

\cite{ma2015statistical} also considered Equation~(\ref{eqn5}), and first emphasized that when Condition~(2) does not hold in practice, the estimation variance of the basic leverage subsampling method, i.e., $\mbox{AVar}(\widetilde{\B}_{BLEV})$, may be inflated to arbitrary large.
Such a case usually happens when there exist some data points whose leverage scores dominate the others, and consequently, there exist some leverage scores that are extremely small.
To combat such a limitation of the basic leverage subsampling method, \cite{ma2015statistical} proposed the shrinkage leverage subsampling method (SLEV), which utilizes the following sampling probability
\begin{equation*}
    \pi_i^{SLEV} = \alpha \frac{h_{ii}}{p} +(1-\alpha)\frac{1}{n}, \quad i=1,\ldots,n.
\end{equation*}
Here, $0<\alpha<1$ is a pre-specified parameter, and it is suggested to choose $\alpha\in[0.8, 0.9]$ in practice \citep{ma2015statistical}.
Intuitively, the SLEV sampling probability $\{\pi_i^{SLEV}\}_{i=1}^n$ interpolates between the basic leverage sampling probability $\{\pi_i^{BLEV}\}_{i=1}^n$ and the equally weighted sampling probability.
Analogous to the BLEV method, SLEV assigns larger sampling probability weights to the data points with high leverage scores.
In addition, SLEV overcomes the disadvantage of BLEV by providing the sampling probability a lower bound, i.e., $(1-\alpha)/n$, which can thus avoid an arbitrary large estimation variance.  

Considering Equation~(\ref{eqn5}), one natural question is whether there exists an optimal sampling probability $\{\pi_i\}_{i=1}^n$ that minimizes the asymptotic variance of the corresponding estimator?
In other words, if given the same tolerance of uncertainty, i.e., to achieve a certain small standard error, which subsample estimator requires the smallest subsample size $r$?
To answer these questions, \cite{ma2020asymptotic} utilized the method of Lagrange multipliers, showing that such an optimal sampling probability exists, which can be written as

\begin{equation*}
\pi_i^{IC} = \frac{||(\X^\T\X)^{-1}\x_i||}{\sum_{i=1}^n ||(\X^\T\X)^{-1}\x_i||}, \quad i=1,\ldots,n.
\end{equation*}
The subsample estimator corresponding to $\{\pi_i^{IC}\}_{i=1}^n$, called the inverse-covariance estimator, has the smallest AMSE that equals
\begin{equation*}
    \mbox{AMSE}(\widetilde{\B}_{IC}) = \sigma^2\mbox{tr}\left\{(\X^\T\X)^{-1}\right\}+\frac{\sigma^2}{r}\sum_{i=1}^n \frac{||(\X^\T\X)^{-1}\x_i||^2}{\pi_i^{IC}}.
\end{equation*}

In addition to making inferences on $\B_0$, researchers may also be interested in estimating $\y = \X\B_0$, since inference on the true regression line $\X\B_0$ is crucially important in regression analysis.
\cite{ma2020asymptotic} proved that the so-called root leverage subsample estimator $\widetilde{\B}_{RL}$ yields the smallest AMSE respecting $\X\B_0$.
Recall that $h_{ii} = \x_i^T(\X^\T \X)^{-1}\x_i$ is the leverage score of the $i$-th data point.
The sampling probability respecting $\widetilde{\B}_{RL}$ can be depicted as
\begin{eqnarray}\label{eqn6}
\pi_i^{RL} = \frac{||\X(\X^\T \X)^{-1}\x_i||}{\sum_{i=1}^n ||\X(\X^\T \X)^{-1}\x_i||} = \frac{\sqrt{h_{ii}}}{\sum_{i=1}^n \sqrt{h_{ii}}}, \quad i=1,\ldots,n,
\end{eqnarray}
where the last equality in Equation~(\ref{eqn6}) is due to the fact that
\begin{equation*}
    ||\X(\X^\T \X)^{-1}\x_i||^2 = (\X(\X^\T \X)^{-1}\x_i)^\T(\X(\X^\T \X)^{-1}\x_i) = \x_i^\T(\X^\T \X)^{-1}\x_i = h_{ii}.
\end{equation*}
Besides, \cite{ma2020asymptotic} also considered the so-called predictor length subsample estimator $\widetilde{\B}_{PL}$ with the sampling probability
\begin{equation*}
\pi_i^{PL} = \frac{||\x_i||}{\sum_{i=1}^n ||\x_i||}, \quad i=1,\ldots,n,
\end{equation*}
and demonstrated that $\widetilde{\B}_{PL}$ has the smallest AMSE respecting $\X^\T\X\B_0$.
Note that when $\X^\T\X = \mathbf{I}_p$, it is easy to check that $\pi_i^{IC} = \pi_i^{RL} =\pi_i^{PL}$, $i=1,\ldots,n$.
The authors refer to \cite{ma2020asymptotic} for a more in-depth comparison for these randomized subsample estimators.

\section{Optimal subsampling methods}

The present section introduces a family of optimal subsampling methods to approximate the full sample least squares estimation. 
Different from the aforementioned randomized subsampling methods, optimal subsampling methods aim to identify a deterministic subsample, whose corresponding subsample estimator is most effective in approximating the true coefficients or the full sample least squares.
In most cases, such a deterministic subsample is selected based on certain rules, especially optimality criteria developed in the design of experiments, e.g., $A$-, $D$- and $E$-optimality \citep{pukelsheim2006optimal}.
An ordinary least squares estimate is then calculated based on the selected subsample.
Some essential background of the optimality criteria is introduced below, followed by some representatives of the optimal subsampling methods.

\subsection{Optimal design techniques and optimality criteria}

As a class of experimental designs, optimal design techniques aim to construct a set of $k$ design points within a given bounded design space, such that these design points are most effective for the estimation of statistical models \citep{pukelsheim2006optimal}.
Under the setting of linear regression models, as shown in Model~(\ref{mmm1}), after the sample is collected, the OLS estimator $\widehat{\B}_{OLS}$ is usually utilized to estimate the true parameter $\B_0$.
The OLS estimator $\widehat{\B}_{OLS}$ is known to be the best linear unbiased estimator, whose mean squared error (MSE) respecting $\B_0$ can be written as $\sigma^2(\X^\T\X)^{-1}$, where $\sigma^2$ is an unknown constant.
The goal of optimal design in the linear regression model is thus equivalent to construct the design matrix $\X$ that minimizes $(\X^\T\X)^{-1}$.

In the cases when the number of predictors $d\geq 2$, $(\X^\T\X)^{-1}$ is a matrix, and is therefore difficult to minimize.
To combat the obstacle, existing optimal design techniques aim to minimize a summary statistic of $(\X^\T\X)^{-1}$, as a surrogate to the matrix $(\X^\T\X)^{-1}$ itself.
In particular, such summary statistics are some real-valued functions that compress a matrix into a real number.
Some popular choices of the summary statistics include the trace of a matrix, the determinant of a matrix, and the maximum eigenvalue of a matrix.
These summary statistics, summarized below, are associated with the $A$-, $D$- and $E$-optimality criteria in optimal designs.

\begin{itemize}
    \item $A$-optimality: aim to minimize the trace of the $(\X^\T\X)^{-1}$.
    \item $D$-optimality: aim to minimize the determinant of $(\X^\T\X)^{-1}$.
    \item $E$-optimality: aim to minimize the maximum eigenvalue of $(\X^\T\X)^{-1}$.
\end{itemize}

\subsection{Subsample selection based on optimality criteria}

Recall that $\{y_i^*, \x_i^*\}_{i=1}^r$ represents a selected subsample.
Consider the following subsample-based least squares estimator
\begin{eqnarray*}
\widetilde{\B}_{\X^*}=(\X^{*\T}\X^*)^{-1}\X^{*\T}\y^*,
\end{eqnarray*}
where $\X^*=(\bm{x}_1^*,\ldots,\bm{x}_r^*)^\T$.
The estimation variance of $\widetilde{\B}_{\X^*}$ respecting $\B_0$, conditional on the sample $\{\x_i\}_{i=1}^n$, can be written as $\sigma^2(\X^{*\T}\X^*)^{-1}$.
Therefore, a natural way to construct the optimal subsample estimator is to find the subsample $\{\x_i^*\}_{i=1}^r$ that minimizes the estimation variance $\sigma^2(\X^{*\T}\X^*)^{-1}$.
Analogous to optimal designs, one can minimize a summary statistics of $(\X^{*\T}\X^*)^{-1}$ as a surrogate to achieve the goal.

Following this line of thinking, an increasing number of methods have been dedicated to selecting the optimal subsample based on a certain optimality criterion in the recent decade.
\cite{wang2019information} considered the $D$-optimality criterion and proposed a heuristic algorithm to select the $D$-optimal subsample.
For each column, the proposed algorithm selects the data points respecting the smallest and the biggest elements of this column.
The final subsample is the one that combines the selected data points respecting all the columns.
\cite{harman2020greedy} proposed two greedy heuristic algorithms for constructing the optimal subsample with respect to the $D$-optimality criterion.
\cite{wang2017computationally} and \cite{nikolov2019proportional} considered selecting the optimal subsample approximately based on the $A$-optimality criterion.
\cite{allen2017near} proposed polynomial-time algorithms for numerous classical optimality criteria, such as $A$-optimality,
$D$-optimality, $V$-optimality, and others.

\subsection{Robust subsample selection}
Although the existing subsampling methods have already shown extraordinary performance on coefficients estimation and model prediction, their performance highly depends on the model specification.
In particular, both the leverage scores and the optimality criteria, introduced in Section~3.1 and Section~4.1, respectively, are derived based on linear regression models.
Nevertheless, the model specification is a trial-and-error process, during which a postulated model could be misspecified. 
The subsampling method derived from a postulated linear regression model does not necessarily lead to a decent subsampling method for the true model.
In other words, the model-based subsampling methods may lead to unacceptable results when the model is misspecified.

In practice, the true underlying model almost always remains unknown to practitioners. 
The subsampling hence is highly desirable to be robust to possible model misspecification. 
To achieve the goal, \cite{meng2020lowcon} considered the setting that the linear regression model is a postulated model, and the true model contains both a linear part and unknown misspecification.
The authors provided an analytic framework for evaluating the AMSE of the subsample least squares estimator in a misspecified linear model.
Recall that when the linear regression model is correctly specified, the least squares estimator is unbiased to true coefficients.
However, when the model is misspecified, \cite{meng2020lowcon} showed that it is extremely easy to construct a "worst-case" subsample, which yields a subsample least squares estimator that may have an arbitrarily large estimation bias.

To combat the obstacle, \cite{meng2020lowcon} aims to select a subsample, which balances the trade-off between bias and variance, in order to yield a robust estimation of coefficients.
To achieve the goal, the authors showed that it is equivalent to select a subsample whose information matrix has a relatively low condition number, a traditional concept in numerical linear algebra \citep{trefethen1997numerical}.
They then proposed an efficient algorithm, called "Lowcon," to identify a subsample with a relatively low condition number. 
In addition, they also showed that the proposed subsample estimator has a finite upper bound of the mean squared error, and it approximately minimizes the "worst-case" bias, with respect to all the possible misspecification terms, under some regularity conditions.

\section{Real data examples}
This section evaluates the performance of different subsampling methods on two real-world datasets, respecting both estimation accuracy and computational time.
The subsampling methods considered here including simple random subsampling (RAND), basic leveraged subsampling (BLEV), shrinkage leverage subsampling (SLEV) with parameter $\alpha=0.9$, root leverage subsampling (RL), inverse covariance subsampling (IC), predictor length subsampling (PL), information-based optimal subset selection (IBOSS) \citep{wang2019information}, Galil-Kiefer method (GKM) \citep{harman2020greedy} and Kumar-Yildirim method (KYM) \citep{harman2020greedy}.
Here, the first six methods are randomized subsampling approaches, summarized in Table~1, while the last three are representatives of the optimal subsampling approaches.

To evaluate the performance of subsampling methods in real data studies, a critical problem is that the true coefficient $\bm{\beta}_0$ is not known to the practitioners.
It is thus impossible to calculate the mean squared error of an estimate respecting the true coefficients.
To overcome this problem, the authors follow \cite{wang2019information}, and use the full-sample ordinary least squares estimator $\widehat{\bm{\beta}}_{OLS}$ as a surrogate for $\bm{\beta}_0$.
In particular, in the $i$-th replicate, $i=1,\ldots, 100$, a bootstrap sample of size $n$ is first uniformly sampled with replacement from the observed sample. 
Next, each subsampling method selects a subsample, resulting in a subsample least squares estimator $\widehat{\bm{\beta}}^{(i)}$.
The subsample sizes are set to be $r=5p, 10p, 15p, 20p$, where $p$ is number of predictors.
The experiments are replicated for one hundred times, and the empirical mean squared error for each subsample estimate is then calculated as
\begin{equation*}
\mbox{EMSE} =\sum_{i=1}^{100}||\widehat{\bm{\beta}}^{(i)}-\widehat{\bm{\beta}}_{OLS}||^2/100.
\end{equation*}
Consequently, an effective subsample estimator should yield a relatively small value of $\mbox{EMSE}$.

\subsection{Example 1: Chemical Sensors Data}
The authors first consider the chemical sensors data, which were collected at the ChemoSignals Laboratory in the BioCircuits Institute, University of California San Diego. 
The dataset contains the readings of 16 chemical sensors exposed to the mixture of Ethylene and CO at varying concentrations in air.
Further information about this dataset can be found in \cite{fonollosa2015reservoir}.

Following \cite{wang2019information}, the authors take the readings from the last sensor as the response and readings from other sensors except for the second one as covariates.
The log-transformation is applied to all the sensors readings since the trace concentrations are often lognormal distributed.
In addition, following \cite{wang2019information}, the first 20,000 data points are excluded since these data points are corresponding to less than 4 min of system run-in time.
The final sample thus contains $n=4,188,261$ observations.
The data are assumed to follow the classical linear regression model, as stated in Model~(\ref{mmm1}).

\begin{table}
\caption{EMSEs of Example 1, with standard deviations presented in parentheses. 
}\label{table2}
\begin{center}
\footnotesize
\resizebox{\textwidth}{!}{
 \begin{tabular}{ l c c c c } 
 \hline
 Method & $r=5p$ & $r=10p$ & $r=15p$ & $r=20p$ \\
 \hline
 RAND & 1.44(0.31) & 0.66(0.36) & 0.43(0.28) & 0.31(0.24) \\   
 BLEV & 0.83(0.37) & $\textbf{0.40}$(0.29) & 0.26(0.27) & $\textbf{0.16}$(0.17) \\   
 SLEV & 1.06(0.42) & 0.47(0.25) & 0.29(0.16) & $\textbf{0.16}$(0.16) \\  
 RL  & $\textbf{0.77}$(0.33) & 0.51(0.23) & $\textbf{0.23}$(0.19) & 0.21(0.16) \\   
 PL & 1.26(0.49) & 0.59(0.29) & 0.38(0.23) & 0.34(0.22)  \\   
 IC & 0.93(0.29) & 0.46(0.26) & 0.28(0.23) & 0.22(0.15) \\   
 IBOSS & 0.91(0.46) & 0.73(0.55) & 0.53(0.35) & 0.18(0.25) \\
 GKM  & 12.30(1.12) & 9.79(0.76) & 10.75(0.87) & 8.81(0.82) \\   
 KYM & 12.18(0.96) & 10.81(1.30) & 11.45(1.25) & 11.52(1.14) \\   
 \hline
 \end{tabular}
}
\end{center}
\end{table}

Table~\ref{table2} summarizes the EMSEs for all the subsample estimates, and the best result in each column is in bold letter.
One can observe that the GKM method and the KYM method perform worse than the RAND method in all cases.
The performance of IBOSS is comparable with the performance of the RAND method.
The randomized subsampling methods, i.e., BLEV, SLEV, RL, PL, and IC, consistently outperforms the RAND method in all the cases.
However, among these five methods, none of them consistently outperforms the others.

The authors then consider the computing time for these subsampling methods when $r=20p$. 
The average CPU time of these methods are $0.001$s(RAND), 3.76s(BLEV), 3.77s(SLEV), 3.81s(RL), 0.49s(PL), 4.49s(IC), 1.20s(IBOSS), 8.59s(GKM) and 2.41s(KYM).
Compared to the CPU time for the full sample OLS estimate, which is 4.62s, all the subsample estimates, except for GKM, requires a shorter time.
The RAND method, as expected, requires the shortest CPU time.
Finally, one can observe that the PL method requires a relatively short CPU time, and it at times yields the best estimation accuracy according to Table~\ref{table2}.

\subsection{Example 2: Diamond Price Prediction}
The second real-world dataset is called \textit{Diamond Price Prediction}, which contains the prices and the features of nearly 54,000 diamonds.
Of interest is to predict the price of the diamond ($y$) by using three continuous predictors ($p=3$): total depth percentage, the weight of the diamond, and width of the top of the diamond relative to widest point. 
The data are assumed to follow the classical linear regression model, as stated in Model~(\ref{mmm1}).

\begin{table}
\caption{EMSEs of Example 2, with standard deviations presented in parentheses. 
}\label{table3}
\begin{center}
\footnotesize
\resizebox{\textwidth}{!}{
 \begin{tabular}{ l c c c c } 
 \hline
 Method & $r=5p$ & $r=10p$ & $r=15p$ & $r=20p$ \\
 \hline
 RAND & 6.02(2.68) & 3.08(1.98) & 1.77(1.51) & 1.52(1.66) \\   
 BLEV & $\mathbf{4.54}$(2.51) & $\mathbf{1.67}$(1.62) & $\mathbf{1.08}$(1.26) & 0.94(1.27) \\   
 SLEV & 5.32(2.96) & 1.96(1.71) & 1.30(1.34) & $\mathbf{0.82}$(1.08) \\  
 RL  & 5.73(2.99) & 1.74(1.49) & 1.33(1.37) & 0.90(1.05) \\   
 PL & 5.32(2.66) & 3.01(2.04) & 1.63(1.71) & 1.41(1.41)  \\   
 IC & 5.45(2.92) & 1.92(1.56) & 1.60(1.27) & 1.00(1.20) \\   
 IBOSS & 9.70(0.57) & 8.80(0.49) & 8.50(0.41) & 8.31(0.43) \\
 GKM  & 11.02(0.53) & 10.89(0.58) & 10.87(0.51) & 10.95(0.52) \\   
 KYM & 10.87(0.66) & 10.92(0.68) & 10.65(0.67) & 10.80(0.65) \\   
 \hline
 \end{tabular}
}
\end{center}
\end{table}

Table~\ref{table3} summarizes the EMSEs for all the subsample estimates, and the best result in each column is in bold letter.
Analogous to Table~\ref{table2}, one can observe that the GKM method and the KYM method perform worse than the RAND method in all cases.
However, different from Table~\ref{table2}, one can observe that IBOSS is not performing well in Table~\ref{table3}.
The authors attribute such an observation to the fact that the second real data example shows less linear pattern than the first one does.
In particular, the authors find that the coefficient of determination, i.e., the $R^2$, for both datasets are equal to 0.999 and 0.854, respectively.
The IBOSS method tends to yield an unpleasant performance when the $R^2$ of the dataset is relatively away from one.
Finally, analogous to Table~\ref{table2}, one can observe that the randomized subsampling methods consistently outperform the RAND method in all the cases.
However, among these randomized subsampling methods, none of them consistently outperforms the others.

The average CPU time for these methods ($r=20p$) are 1.20ms (RAND), 5.00ms (BLEV), 5.53ms (SLEV), 6.40ms (RL), 3.53ms (PL), 6.30ms (IC), 3.85ms (IBOSS), 5.70ms (GKM), and 2.40ms (KYM).
Analogous to the first example, most of these methods require a shorter CPU time than the full sample OLS estimate (6.4ms).
One can also observe that the RL method has a decent estimation accuracy and requires a relatively short CPU time, thus takes the balance between the estimation accuracy and the computational cost.

\section{Conclusion and future research}
This paper reviews modern subsampling methods for solving large-scale least squares regression problems, focusing on the randomized subsampling methods and the optimal subsampling methods.
The authors discuss the theoretical advantages of these methods over the simple random subsampling method.
Real data studies show most of the existing subsampling methods approximate the least squares estimate effectively and efficiently.
Among these methods, it is observed that the predictor length subsampling method tends to take the balance between the estimation accuracy and the computational cost most effectively.

Most reported research efforts are exploring subsampling methods for least squares problems. In contrast, there is relatively little reported work in literature discussing subsampling methods for other models or algorithms such as generalized linear models, nonparametric regression, kernel methods, time series models, variable selections, and deep learning. 
This happens due to the fact that these models have a more complicated formulation than linear models.
Thus, the optimal sampling probability (or the optimality criteria) and efficient subsampling algorithms respecting these models may be hard to derive. 
For example, \cite{wang2018optimal} proved that the leverage scores for the logistic regression model are dependent on the response variable, thus resulting in a more complicated leverage-based subsampling algorithm than the one for the linear regression model.
More research efforts are needed to develop effective and efficient subsampling algorithms beyond least squares estimation.

Another interesting research issue open for future investigation is to develop robust subsampling methods. 
Most of the existing subsampling methods require that the model is correctly specified. 
Practically, however, the true underlying model is almost always unknown to practitioners.
The subsampling method derived from a postulated model may generate unacceptable results when the model is misspecified.
Thus, how to improve the robustness of subsampling methods is an essential topic for future investigation.

\newpage
\bibliographystyle{chicago}
\bibliography{main}

\end{document}